\documentclass[10pt,letterpaper]{article}

\usepackage[T1]{fontenc}
\usepackage[utf8]{inputenc}
\usepackage{graphicx, amssymb, bm, eucal, float}
\usepackage[square,sort,comma,numbers]{natbib}

\def\a{\alpha}
\def\b{\beta}

\def\f{\phi} \def\vf{\varphi}
\def\g{\gamma}
\def\h{\eta}

\def\l{\lambda}
\def\m{\mu}

\def\r{\rho}
\def\s{\sigma}

\def\L{\Lambda}

\def\fr{\frac}

\def\pro{\propto}
\def\la{\left}
\def\ra{\right}
\def\pa{\partial}

\def\inf{\infty}

\def\bar#1{\overline{#1}}

\def\ba{\begin{array}}
\def\ea{\end{array}}
\def\be{\begin{equation}}
\def\ee{\end{equation}}
\def\bdm{\begin{displaymath}}
\def\edm{\end{displaymath}}
\def\bea{\begin{eqnarray}}
\def\eea{\end{eqnarray}}
\def\nl{\nonumber \\}

\def\lb{\label}
\def\sp{~~~}

\begin{document}

\title{From Cosmic Inflation and Matter Creation to\\Dark Matter - Journey of the Inflaton?}
\author{B. S. Balakrishna\footnote{Email: balakbs2@gmail.com}}
\date{December 12, 2022 \\ Revised: April 6, 2023}

\maketitle

\begin{abstract}

A scenario of the inflaton evolution from cosmic inflation and matter creation to dark matter/dark energy today is presented. To start with, a model of the inflationary phase of the inflaton is introduced. The inflaton rolls down a hilltop potential along with matter creation being dragged down by the presence of matter. Presence of matter provides a mechanism to stop universe's acceleration and hence the inflationary phase. The model predictions for the standard metrics are fully consistent with the current CMB limits. The potential could in principle be extended to complete a potential hill subsequent to inflation. The evolution of the inflaton from the inflationary phase to radiation/matter dominated eras and to current times can be inferred qualitatively following the evolution of its equation of state parameter. Existence of solutions to its dynamics, tracking matter as it evolves to current times, provides a plausible reasoning for the relative order of magnitudes of the cosmological parameters, in particular to the relative abundance of dark matter today.

\end{abstract}

\section{Introduction}

Recent observations have confirmed the presence of dark matter/dark energy in the universe today. Within the framework of the $\L$CDM model of the universe, the data indicates about 34\% matter and 66\% dark energy. Because observations reveal only about 5\% ordinary matter, the rest 29\% has been attributed to cold dark matter, leading to dark matter searches. In the absence of observational evidence for dark matter, various alternatives have been suggested, including the possibility that a scalar background could be the source of dark matter/dark energy.

In \citep{bs1}, the plausibility that a scalar background could be the source of dark matter/dark energy is explored within the context of solvable solutions to the governing Friedmann–Lema\^itre–Robertson–Walker (FLRW) equations. It was shown that such solvable scalar potentials are consistent with the combined dynamics in such a universe, with certain potentials providing satisfactory fits to Type Ia supernovae (SNe Ia) data. The kinetic and potential energies of the scalar provided the source for dark matter and dark energy with the scalar rolling down the potential as the universe expands. Similar scenarios with the potential playing the role of dark energy or a time-varying cosmological constant have been discussed lately in the literature under the name of quintessence\citep{rpx,cds,dxx}. 

An appealing follow up on the above approach is to explore the possibility that the scalar source may be a remnant of cosmic inflation. Such an identification has been pursued earlier in the literature\citep{sxx, pvx, prx, dvx, wxx}. Here in the article, this approach is further explored presenting a scenario of the inflaton evolution from cosmic inflation to current times. To provide a consolidated picture, a model of the inflationary phase of the inflaton is presented that is fully consistent with the current CMB limits on the standard metrics. In the model, the inflaton rolls down a hilltop potential along with matter creation, slowly and naturally, being dragged down by the presence of matter. Matter creation during inflation can have many advantages, in particular its presence can have a dramatic impact as to how long inflation continues. Such models provide a natural mechanism to stop universe's acceleration and hence the inflationary phase, while the inflaton is still on a slow-roll down the potential hill.

The potential obtained in the inflationary phase could in principle be extended with the potentials implied from SNe Ia data to complete the potential hill governing inflaton dynamics. The evolution of inflaton's equation of state parameter subsequent to inflation provides a qualitative picture of the inflaton's evolution, to radiation/matter dominated eras and to current times. An intriguing aspect of certain scalar models in a FLRW universe is the existence of tracking solutions, as has been noted earlier in \cite{clw} within the context of exponential potentials. Existence of tracking solutions enables inflaton to behave as cold dark matter during universe's expansion, with dark energy further arising as a result of the slowing down of the inflaton due to `Hubble friction' as it evolves to current times. Such a likely scenario of the inflaton's evolution provides a plausible reasoning for the relative order of magnitudes of the cosmological parameters, in particular that of dark matter abundance today.

\section{FLRW Equations}
\lb{flr}

Let us consider a homogeneous, isotropic and spatially flat universe in the presence of matter and a single scalar field that we identify as the inflaton. Normal matter is referred to as `matter' with no qualifier. Scalar field is taken to be spatially uniform but time-varying. In this universe, FLRW equations for the scale factor $a(t)$ read (in units $c=1$ and $8\pi G=1$)
\bea
3\fr{\dot{a}^2}{a^2} &=& U+V+\r, \nl
2\fr{\ddot{a}}{a}+\fr{\dot{a}^2}{a^2} &=& V-U-p.
\lb{aeq}
\eea
A dot on a symbol denotes time differentiation. A prime on a symbol will denote differentiation with respect to $a$ in what follows. Matter density is $\r$ and its pressure is $p$. $U$ is the kinetic energy and $V$ is the potential energy of the scalar field $\f$, such that
\be
V = V(\f), \sp U = \fr{1}{2}\dot{\f}^2 = \fr{1}{2}\dot{a}^2\f'^2.
\ee
Scale factor is chosen to be unity at present time. It is convenient to introduce $f(a)$ for the inflaton energy density, called `fit-function' in \citep{bs1}, defined as
\be
f = U + V.
\ee
FLRW equations for evolution with respect to $a$ can now be equivalently written as
\bea
V &=& f+\fr{a}{6}f', \nl
a^2\f'^2 &=& \fr{-af'}{f+\r}.
\lb{vf2}
\eea
Note that for $U=f-V$ we just have
\be
U = -\fr{a}{6}f'.
\ee
It is assumed that matter-side of dynamics is governed by the fluid equation in an adiabatic evolution of the universe,
\be
a\r'+3(\r+p) = 0.
\lb{mdn}
\ee
In the presence of matter creation, $p$ would include creation pressure in addition to `true' pressure\citep{pgn}. In such a case, for our purpose, above could be considered as determining pressure itself. In the absence of matter creation, given the equation of state $p=w\r$, co-moving density $\s$ is defined as
\be
\s = \r a^{3(1+\bar{w})}, \sp \bar{w}=\fr{1}{{\rm ln}a}\int_1^a\fr{da}{a}w(a).
\lb{six}
\ee
For constant $w$, we just have $\bar{w}=w$. In terms of $\s$, matter dynamics (\ref{mdn}) simplifies to $\s'=0$, that is, $\s$ is a constant over time. As for the the equation of state parameter of the scalar $w_s$ itself, it is expressible as
\be
w_s = \fr{U-V}{U+V} = -\fr{1}{3a^2f}\la(a^3f\ra)' = -1-\fr{a}{3f}f' = -1+\fr{1}{3f}(f+\r)a^2\f'^2.
\lb{eqs}
\ee
This can cover a wide range within $[-1,1)$ depending on the potential, for instance $w\to 1$ if $V/f\to 0$. This indicates that the scalar can evolve effectively as radiation or matter over a period of time.

Given a fit-function $f(a)$, results for $V$ and $\f'$ provide a relation between them in terms of `parameter' $a$. This could be used to imply $V(\f)$ as a function of $\f$, first solving for $\f$ in terms of $a$ and then using $a$ as a function of $\f$ in the expression for $V(\f(a))$. As noted in (\cite{bs1}), such solvable potentials provide a self-consistent framework, with the equation of motion of $\f$ being automatically satisfied. Specifically, FLRW equations imply (assuming $\f'\ne 0$)
\be
\fr{1}{a^3}\fr{d}{dt}\la(a^3\fr{d\f}{dt}\ra)+\fr{dV}{d\f}+\fr{1}{a\f'}\la(a\r'+3(\r+p)\ra) = 0.
\lb{fre}
\ee
If the matter dynamics is independently satisfied as in Eq. (\ref{mdn}) (or $\s'=0$), the second term vanishes, and we get the equation of motion satisfied in the $a-$background.

Results for $V$ and $\f'$ are not tractable in general given a potential $V(\f)$. One could simplify them by introducing an intermediate function $u(\f)$ defined as
\bea
u(\f(a)) &=& \int^adb~b\f'^2(b), \nl
u_{\f} = \fr{du(\f)}{d\f} &=& a\f'(a), \sp a \pro {\rm exp}\la(\int^{\f}\fr{d\f}{u_{\f}}\ra).
\eea
Integration base limit could be suitably chosen, for instance as zero, one or $\inf$. In terms of $u(\f)$, we can express $f$, given some constant $C$, as
\be
f = Ce^{-u(\f)} + \int_a\fr{db}{b}\r(b)u_{\vf}^2e^{u(\vf)-u(\f)},
\lb{tfb}
\ee
where $\f=\f(a)$ and $\vf=\vf(b)$. The potential can now be expressed as
\be
V(\f) = \la(1-\fr{1}{6}u_{\f}^2\ra)f-\fr{1}{6}u_{\f}^2\r.
\ee
We thus get a nonlinear integro-differential equation for $u(\f)$. Analytic solutions are not available in general, and one may need to resort to numerical computations. 

\section{Inflation with Matter Creation}
\lb{imc}

It is often argued that matter, if any, will be swept away during inflation. As a consequence, it becomes necessary to introduce a subsequent process called reheating. Here let us explore the possibility that matter is created during inflation itself, and with sufficient intensity so that it survives the inflationary tide. Besides creating matter for the universe, this has other significant advantages. Presence of matter provides a natural drag on inflaton ensuring slow-roll down the potential. More importantly, it provides a mechanism to stop inflation while still on slow-roll. Furthermore, it sets the normal matter to dark matter ratio at early times to let it further evolve to current values as the universe expands.

It is convenient to consider $f(a)$ as a function of field $\f$ in the following, so that $af'=a\f'f_{\f}$ where a subscript ${\f}$ on a symbol denotes differentiation with respect to $\f$. Given this, FLRW equations (\ref{vf2}) can be expressed as
\be
V = f-\fr{1}{6(f+\r)}f_{\f}^2, \sp a\f' = -\fr{1}{f+\r}f_{\f}.
\lb{vff}
\ee
This is a nonlinear first order differential equation to be solved for $f_{\f}$, and hence $\f(a)$, given a potential $V(\f)$ and a model of matter density $\r$. The inflaton, perched near the top of a hilltop potential, will start slowly rolling down the potential according to second equation above, initiating the inflationary phase of its evolution. For inflation to occur, we require that universe accelerates, that is
\be
\ddot{a} = \fr{1}{6}(a^2(f+\r))' > 0.
\lb{acc}
\ee
If matter is created during inflation itself, rising $\r$ could potentially compensate falling $f$ such that the above is easily satisfied, and inflation will continue. In the absence of matter creation above reads
\be
a^2\f'^2 < 2-\fr{3(1+w)\r}{f+\r} \sp {\rm or} \sp w_s < -\fr{1}{3}-(1+3w)\fr{\r}{3f},
\lb{af2}
\ee
where $w$ is the equation of state parameter of matter. If this fails to satisfy, we may assume that matter creation could not be supported, and the universe would stop accelerating. With a significant matter creation rate, this inequality would get first violated when $a^2\f'^2$ remains small, or at least not too close to 2, resulting in $\r/f$ of order one. If $a^2\f'^2$ remains small (or $w_s\sim-1$), the inequality will get first violated when
\be
\r \simeq \fr{2f}{(1+3w)} = f.
\ee
In the last step above and in the following, we assume radiative matter ($w=1/3$). In addition to ending inflation, this sets the normal matter to inflaton density ratio at early times, to let it further evolve to current values as the universe expands. With matter subsequently evolving as $a^{-3(1+w)}$, the inequality may get hit again, but this does not last more than a couple of e-folds due to subsequent rapid rise of $a^2\f'^2$ (result of falling $\r/f$ and an expected tracking solution picked up by the evolution).

One could model a hilltop potential in multiple ways near its top. For instance, one could assume that it is quadratic in $\f$ at the top say near $\f=0$. If so, we could consider solving it perturbatively so that, to $\mathcal{O}(\f^2)$, we have
\be
V(\f) = A-\fr{1}{2}\m\f^2, \sp f = A-\fr{1}{2}\l C\f^2, \sp \r = C-A+\fr{1}{2}\g C\f^2,
\ee
where $\m,\l,\g$ are constant positive parameters, and $C$ can be viewed as the initial total energy density, that is $f+\r$ at $\f=0$. Inserting these into Eq. (\ref{vff}), we get to $\mathcal{O}(\f^2)$
\be
a\f' = \l\f, \sp \f = \f_0(a/a_0)^{\l}, \sp \m = \l\la(1+\fr{\l}{3}\ra)C.
\ee
The inflaton, starting at $a=a_0$ from some point $\f=\f_0>0$ infinitesimally small, rolls down the potential away from the hilltop. One may verify that the above solution does satisfy the equation of motion for the inflaton in the $a$-background. For the matter density, we have
\bea
\r &=& \r_0+\fr{1}{2}\g C\la(\f^2-\f_0^2\ra), \sp \r_0 = C-A+\fr{1}{2}\g C\f_0^2, \nl
a\r' &=& \l\g C\f^2 = 2\l(\r-\r_0)+\l\g C\f_0^2.
\eea
For the inflaton starting at $\f_0=0$ with $\r_0=0$, we have $a\r'=2\l\r$. Interestingly, we can obtain some values for the parameters. The scalar spectral index to $\mathcal{O}(\f^2)$ is given by (approximating $k\pa_k\sim a\pa_a$ where $k\sim aH$ is the Fourier vector and $H$ the Hubble parameter),
\be
1-n_s \simeq a\la({\rm ln}(a^2\f'^2H^{-2})\ra)' \simeq 2a\la({\rm ln}(a\f')\ra)' = 2a\la({\rm ln}\f\ra)' = 2\l.
\lb{ns1}
\ee
Variation of $H$ leads to a higher order correction $\pro (\g-\l)\l\f^2$. CMB observation $n_s\simeq 0.96$ tells us that $\l\simeq 0.02$. This being a constant, scalar spectral index has no scale dependence to $\mathcal{O}(\f^2)$ and all its running indexes will be nearly zero consistent with the current CMB observations.

Given our solution $f+\r\pro 1+(\g-\l)\f^2/2$, inequality (\ref{acc}) is easily satisfied for $\g\sim\l$ and inflation will continue until say $a=a_e$, $\f=\f_e$ where
\be
\f_e^2 = (2-\h)\f_0^2+\fr{\h}{\l}\la(1-\fr{2\r_0}{C}\ra), \sp \h = \fr{2\l}{\g+\l}.
\ee
For $\g\sim\l$, we have $\h\sim 1$. If inflaton started off right from the top of the potential with $\r_0=0$, we get $\f_e^2=\h/\l\simeq 50\h$. Given that $a^2\f_e'^2=\l^2\f_e^2\simeq 0.02\h$ is relatively negligible, inflation would end with $\r\sim f$ as noted above. For the number of e-folds from a point $\f$ to end of inflation, we have ($\f_0=0$, $\r_0=0$)
\be
N = {\rm ln}\la(\fr{a_e}{a}\ra) = \fr{1}{2\l}{\rm ln}\la(\fr{\f_e^2}{\f^2}\ra) = \fr{1}{2\l}{\rm ln}\la(\fr{\h}{\l\f^2}\ra) ~>~ 53 ~~{\rm if}~~ \f^2<6\h.
\ee
Let us look at some other parameters, namely the tensor to scalar ratio and the tensor spectral index. For the tensor to scalar ratio, we have
\be
r = 8a^2\f'^2 = 8\l^2\f^2 \simeq 0.0032\f^2.
\ee
For $N\simeq 55$ e-folds relevant for today's horizon, corresponding to pivoting at say $a=a_p$ or $\f=\f_p$, we get $\f_p^2\simeq 5.5\h,~r\simeq 0.018\h$. The model does not satisfy the usual consistency relation $n_t=-r/8$ where $n_t$ is the tensor spectral index. Instead, we have
\be
n_t \simeq a\la({\rm ln}(H^2)\ra)' = a\la({\rm ln}(f+\r)\ra)' = (\g-\l)\l\f^2 = \fr{1}{4}\la((1/\h)-1\ra)r.
\ee
For the power spectrum amplitude at $a=a_p$,
\be
P_S = \fr{H^2}{4\pi^2a_p^2\f_p'^2} = \fr{H^2}{4\pi^2\l^2\f_p^2} \sim 10H^2.
\ee
Given the CMB data $P_S\sim 10^{-9}$, we get $C\sim H^2\sim 10^{-10}\sim (10^{16}{\rm GeV})^4$.

Density level matter creation rate $\g=\l$ appears special, resulting in the total energy density $f+\r$ held constant during inflation. This is an interesting case since the constancy implies a perfect flat de Sitter universe, but in the presence of inflaton and matter. Constancy case would make it an adiabatic evolution within the horizon, supplementing the general case interpretation within a co-moving volume of the adiabaticity of the inflaton's equation of motion and the fluid equation of matter.

Presence of matter provided a natural drag on the inflaton's evolution resulting in its slow-roll down the potential. In the standard hilltop potential scenario where matter is absent, the inflaton would be rolling down relatively faster, especially when nearing the end of inflation. This follows from the rolling speed $a'\f'=-af_{\f}/f$ in such a case compared to $a'\f'=-af_{\f}/(f+\r)$ as we have in the present model. Further, in such scenarios, the rolling speed is expected to reach $\sqrt{2}$ to stop inflation, while it is expected to be relatively small in the present model. However, matter creation is considered as part of the inflationary phase. Inflation being an accelerated phase, this is very much a possibility. In fact, gravitationally induced matter creation has been a subject of much interest in the literature (see for instance \cite{fxx}). An appealing possibility is that matter fluctuations arose out of curvature fluctuations during inflation, and survived to live beyond inflation in analogous fashion to inflaton fluctuations. This could provide a rationale for our assumption that matter creation was active only during inflation and ended rather abruptly when the universe stopped accelerating. Curvature fluctuations, regarded as a superposition of inflaton and matter contributions under first order perturbation, would evolve with the inflaton component becoming dominant by the time of last scattering, justifying our use of the inflaton driven expression in the scalar spectral index in Eq. (\ref{ns1}).

The potential, assumed to be quadratic all the way up to end of inflation, could be subject to corrections. Depending on the full potential of the inflaton, there could be higher order terms that may become relevant, especially near end of inflation. If necessary, one could look for corrections to our results perturbatively, solving Eq. (\ref{vff}) say for a quartic term in the potential, assuming constancy of $f+\r$ as a model of matter creation. This would add a $\f^3$ correction to $a\f'$ and hence a $\f^2$ correction to the scalar spectral index. Such corrections are in any case expected beyond inflation to make the potential bounded from below.

As noted earlier, one could model a hilltop potential in multiple ways near its top. Besides the quadratic class of hilltop potentials, a class of potentials discussed in the literature are those with exponential tails defining a nearly flat hilltop plateau for large negative values of $\f$. An example of such a potential is $1/(1+e^{\b\f})$ for some positive constant $\b$. Such a potential can be analyzed analogously, and is presented elsewhere within the context of a numerical evolution of the inflaton field.

\section{Inflaton As Dark Matter}

Function $f(a)$, that we referred to as the `fit-function', can be conveniently modeled to fit SNe Ia data. This was done in \citep{bs1} for various choices that provided satisfactory fits. The simplest one tried was the $\L$CDM fit itself:
\be
f(a) = \a_0+\fr{\a_1}{a^3}, \sp \a_0 = 0.66\r_c, ~ \a_1+\r_1 = 0.34\r_c,
\lb{mdl1}
\ee
where $\r_c$ is the current critical density and $\r_1$ is the current matter density. $\L$CDM is the limiting case of this model as $\a_1\to 0$, or equivalently as $\r_1\to 0.34\r_c$. Being exactly the one that $\L$CDM generates, it fit the data just as well, but with the dark matter/dark energy supplied by the kinetic and potential energies of the scalar. However, in the model, dark energy can be viewed as the cosmological constant in disguise. But, SNe Ia data can also be fit reasonably well with other fit-functions having no such constants. For instance, a fit mimicking flat-wCDM is
\be
f(a) = \fr{\a_0}{a^{\b}}+\fr{\a_1}{a^3}, \sp \a_0 = 0.70\r_c, ~ \a_1+\r_1 = 0.30\r_c, ~ \b=0.5.
\lb{mdl2}
\ee
Flat-wCDM is the limiting case of this model as $\a_1\to 0$, or equivalently as $\r_1\to 0.30\r_c$. In our case, the fit accommodates any matter density up to 30\%. If we choose it to be say 5\% of the critical density to be close to observations, the remaining 25\% is supplied as dark matter by the kinetic and potential energies of the scalar, along with the 70\% as dark energy. In this model, the scalar equation of state parameter $w_s$ is
\be
w_s = -(1-\b/3)\fr{\a_0a^{3-\b}}{\a_0a^{3-\b}+\a_1}.
\ee
It has the value $\simeq -0.6$ today and will tend towards $\simeq -0.8$ in the future. These numbers are model dependent, but provide the kind of values to expect. But, more interestingly, looking back in time, it has journeyed starting from $w_s\simeq 0$ at last scattering to what it is today. In other words, at those early times, as can also be noted directly from (\ref{mdl1}) or (\ref{mdl2}), dark energy term was relatively negligible and the scalar contributed dominantly as cold dark matter to universe's expansion.

All such models providing reasonable fits indicate that $w_s$ is close to zero at those early times. Because the data fits are expected to hold well back in time only up to last scattering, it doesn't appear, given the present data, that we could infer the road $w_s$ took further back in time to reach a value near $0$. If we are to identify this scalar source of dark matter/dark energy with the inflaton responsible for cosmic inflation, then we could speculate on the road the inflaton may have taken in reaching a near zero value around last scattering. Interestingly, this is the opposite of what the scalar has been taking at recent times.

As noted in the last section, inflation ended when inflaton energy density and matter density are about the same, that is $f\sim\r$. Matter creation is expected to have ended subsequently following which $f$ and $\r$ would decay (redshift under scale factor). Their decay rates are related to their respective equation of state parameters $w_s$ and $w$. At the end of inflation, inflaton was largely dark energy, or $w_s$ was close to $-1$, as can be noted from (\ref{af2}) given that $f\sim\r$. However, $w_s$ is expected to keep rising at a rate related to rate of increase of $U/V$ ($U=f-V$ being the inflaton kinetic energy) as inflaton speeds up rolling down the potential. $a^2\f'^2$ will follow $w_s$ helped with decreasing $\r/f$ (as long as $w_s<w$). This follows from
\be
w_s' = \fr{2V^2}{(U+V)^2}\la(\fr{U}{V}\ra)', \sp a^2\f'^2 = \fr{3(1+w_s)}{(1+\r/f)}.
\ee
How far $w_s$ and $a^2\f'^2$ will go depends on the potential characteristics. Local expansion of the hilltop potential is not expected to be valid any more. As the universe evolves through subsequent phases, $w_s$ heads towards $w$ with $f$ becoming less of dark energy and more of dark matter. Because $\r/f$ decays at rate $3(w-w_s)$ (as a function of ${\rm ln}a$) that could range from about 4 down to zero as $w_s$ heads towards $w$, one could expect $\r/f$ to stabilize in a few e-folds at least temporarily. In other words, a tracking solution is likely to become dominant with $f$ effectively tracking matter. That such a tracking solution is likely is evident from the data fits like (\ref{mdl2}), and is illustrated later below in the specific case of an exponential potential. If $a=a_p$ and $w_s=w_p$ is such a point in time, we get from Eqs. (\ref{vf2}) and (\ref{eqs})
\be
f_p = \fr{\r_p}{3(1+w_p)/(a_p^2\f_p'^2)-1}, \sp V_p = \fr{1}{2}(1-w_p)f_p.
\lb{imp}
\ee
If $w_p=w$, tracking could be considered perfect and all of inflaton energy would behave as dark matter, decaying just as matter would. However, it is more likely that $w_p<w$ implying the presence of dark energy. The relative insignificance of dark energy relative to dark matter until late in the evolution indicates that $w_p\sim<w$. Hence, above would closely relate to the amount of dark matter in the universe. Since our $w$ is associated with just the normal matter, transition $w=1/3\to 0$ would happen quite late in the evolution. Choosing $w_p\simeq w\simeq 0$ at that point in time and $a_p^2\f_p'^2=2.5$ as an example gives us $f_p\simeq 5\r_p$. If dark matter tracking continues subsequently, this could equate to dark matter abundance at current times.

To illustrate that a tracking solution is likely, let us look for a solution $f\pro \r=\s a^{-3}$ (choosing $w=0$) setting $a^2\f'^2$ to a constant $\b<3$, so that for some $\a$,
\be
af' = -\b(f+\r), \sp \f = \a+\sqrt{\b}~{\rm ln}a.
\ee
This gives the following solution given some constant $C$,
\be
f(a) = Ca^{-\b}+\fr{\b\s}{3-\b}a^{-3}.
\lb{xpx}
\ee
This follows from the potential
\be
V(\f) = Ce^{-\sqrt{\b}(\f-\a)}+\fr{\b\s}{3-\b}e^{-(3/\sqrt{\b})(\f-\a)}.
\ee
The second term in $f(a)$ would correspond to dark matter tracking normal matter. The first term would correspond to dark energy, relatively insignificant at early times, but becoming significant later. As in the previous example, with $\b=2.5$, $w=0$, we get $5\r$ for the second term. The fit function is of the form of a flat-wCDM, but the exponent $\b$ for the first term is larger than expected, failing to provide a good fit to data. But, the assumption that $a^2\f'^2$ is a constant is not expected to hold good either all the way to current times. In fact, as models (\ref{mdl1}) and (\ref{mdl2}) suggest, it is expected to decrease to end up $\sim 1$ today. Hence for a full analysis, its variation needs to be taken into account that can be modeled in various ways. For instance, the above potential with an exponent zero or near zero in the first exponential can be shown numerically to provide a reasonable fit to SNe Ia data.

Matching of the solution to Eq. (\ref{imp}) indicates that $a^2\f'^2$ at early times was about the fraction of dark matter to total matter in the universe (times $\sim 3(1+w)$). This is consistent with the inflaton picture since, having passed through inflation and subsequent phases, $a^2\f'^2<3(1+w_s)$ reached $\simeq 2.5$ around the time of last scattering. Models in \citep{bs1} are expected to address the potential back in time only until about last scattering. Evaluating $a^2\f'^2$ in those models and taking the $a\to 0$ limit should give us its approximate implied value at those times. Models (\ref{mdl1}) and (\ref{mdl2}) are consistent with this picture (ignoring a small effect of radiation on our $w$ at last scattering).

With dark energy becoming more dominant at recent times, tracking is expected to become less effective as a whole. But a component of $f(a)$ as in the solution above  could still be tracking matter contributing as dark matter to the universe's expansion. Looking at the general solution (\ref{tfb}) for $f(a)$, we note that it has a similar decomposition into two components, with the second one that could effectively act as the tracking component to the evolution. It could be recast as $a^{-3}g(a)$ where $g(a)$ can be varying relatively slowly with respect to $a$ for suitably chosen potentials.

Given a picture of the potential hill as the right-half of a bell-shaped curve, it is possible to make some later times observations regarding the inflaton journey. Inflaton sped up at early times rolling down the hill resulting in an increasing $w_s$ until some time before last scattering. It is likely to be on the potential tail ever since, being slowed down by the `Hubble friction' applied to its move by the universe's expansion according to $U=-af'/6\sim-aV'/6$. This and later time solutions such as (\ref{xpx}) suggest that $f(a)$ is also expected to follow a bell-shaped curve. As Eq. (\ref{eqs}) indicates, rate of increase of $w_s$ is proportional to concavity of ${\rm ln}f(a)$ as a function of ${\rm ln}a$. Hence, for such a class of fit-functions or potentials leading to decreasing $U/V$, we thus expect $w_s$ to have reversed its direction and to have started decreasing. As a result $\r/f$, decaying at rate $3(w-w_s)$ (as a function of ${\rm ln}a$), would drop further from its value of $\simeq 1/5$ to what could well be $\simeq 1/19$ today. This is consistent with the tracking picture discussed above since the tracking component would still account for a dark matter abundance of $\simeq 5\r$. With $f\simeq 19\r$, the remaining $14\r$ would contribute as dark energy dominant at recent times.

\section{Conclusions}

Many models of cosmic inflation usually sidestep the question of how long inflation continues, leaving it to the parameters to go out of the handled range. As shown here in the article, presence of matter can provide a natural and dramatic mechanism to stop inflation, while the inflaton is still on a slow-roll down the potential hill. This has an additional advantage of setting the normal matter to dark matter ratio at early times, to let it further evolve to current values as the universe expands. Moreover, given that matter creation is likely to be an irreversible process in an accelerating universe, inflation would cease at all points in due time. If this holds, it would help avoid eternal inflationary scenarios, and also suppress subsequent inflationary bubbles to ensure stability of the universe we live in. However, for these advantages to follow, matter creation needs to be a part of the inflationary phase. Inflation being an accelerated phase of the universe's expansion, this is very much a possibility.

It thus appears possible to have a consistent picture of a single scalar field taking the role of the inflaton responsible for cosmic inflation at early times as well as that of a scalar source of dark matter and dark energy dominant at later times. With the inflaton potential acting effectively as a running cosmological constant in line with such considerations in the literature, it is capable of providing an explanation for the order of magnitudes of the cosmological parameters we observe today. So, if such a picture is realizable, one may rephrase the question of why the `cosmological constant' as implied by the $\L$CDM model is so tiny today by, perhaps a less mysterious but equally puzzling one of, why the inflaton potential doesn't appear to have a relatively large `true' cosmological constant bundled into it as it appears to tend to zero or near zero asymptotically. However, for a full picture of its evolution, one would need to extend the hilltop potential effective at early times to later time potential candidates. Such an investigation of the inflaton's evolution from early times to present day can be attempted numerically thanks to partial tractability of its governing equations. It is presented elsewhere within the context of a potential with exponential tails defining a nearly flat hilltop plateau. For a fully realistic model though, this requires choosing appropriate potential candidates and a better understanding of the potential at intermediate times, that is in the radiation dominated and matter dominated eras.

\newpage

\end{document}